\documentclass[twocolumn,aps,prl,showpacs,superscriptaddress,final]{revtex4-1}   
\usepackage{bm}
\usepackage{hyperref}
\hypersetup{
    colorlinks = true,
    linkcolor  = blue,
    citecolor  = blue,
    filecolor  = magenta,
    urlcolor   = cyan
}
\usepackage [english]{babel}
\usepackage [utf8]{inputenc}
\usepackage [T1]{fontenc}
\usepackage{graphicx}
\usepackage{amsmath}
\usepackage{amssymb}
\usepackage{siunitx}
\usepackage{booktabs}
\usepackage{float}
\usepackage{array}
\usepackage[normalem]{ulem}
\usepackage[dvipsnames]{xcolor}
\definecolor{royalblue}{HTML}{4169e1}

\definecolor{bond}{HTML}{007007}

\providecommand{\ifhighlighting}{\iffalse} 
\ifhighlighting
\usepackage{xcolor}
\newcommand{\hlstart}{\color{purple}}
\newcommand{\hlend}{\color{black}}
\else
\newcommand{\hlstart}{}
\newcommand{\hlend}{}
\fi

\usepackage{booktabs,makecell, multirow, booktabs}

\newcommand{\be}{\begin{equation}}
\newcommand{\e}{\end{equation}}

\newcommand{\beml}{\begin{subequations}}
\newcommand{\eml}{\end{subequations}}
\newcommand{\beq}{\begin{eqnarray}}
\newcommand{\eq}{\end{eqnarray}}
\newcommand{\ba}{\begin{array}}
\newcommand{\ea}{\end{array}}

\usepackage{braket}

\begin{document}
\date{\today}

\title{Universal Crosstalk of Twisted Light in Random Media}

\author{David Bachmann}

\email{david.bachmann@physik.uni-freiburg.de}

\affiliation{Physikalisches Institut, Albert-Ludwigs-Universit\"at Freiburg, Hermann-Herder-Str.\ 3,
D-79104 Freiburg, Germany}

\author{Asher Klug}

\affiliation{School of Physics, University of the Witwatersrand, Private Bag 3, Johannesburg 2050, South Africa}

\author{Mathieu Isoard}

\thanks{Present address: Laboratoire Kastler Brossel, Sorbonne Université, ENS-Université PSL, Collège de France, CNRS, 4 place Jussieu, F-75252 Paris, France.}

\affiliation{Physikalisches Institut, Albert-Ludwigs-Universit\"at Freiburg, Hermann-Herder-Str.\ 3,
D-79104 Freiburg, Germany}

\author{Vyacheslav~Shatokhin}

\affiliation{Physikalisches Institut, Albert-Ludwigs-Universit\"at Freiburg, Hermann-Herder-Str.\ 3,
D-79104 Freiburg, Germany}

\affiliation{EUCOR Centre for Quantum Science and Quantum Computing, Albert-Ludwigs-Universit\"at Freiburg, Hermann-Herder-Str.3, D-79104 Freiburg, Germany}

\author{Giacomo Sorelli}

\thanks{Present address: Fraunhofer IOSB, Ettlingen, Fraunhofer Institute of Optronics, System Technologies and Image Exploitation, Gutleuthausstr. 1, D-76275 Ettlingen, Germany.}

\affiliation{Laboratoire Kastler Brossel, Sorbonne Université, ENS-Université PSL, 
	Collège de France, CNRS; 4 place Jussieu, F-75252 Paris, France}

\author{Andreas Buchleitner}

\affiliation{Physikalisches Institut, Albert-Ludwigs-Universit\"at Freiburg, Hermann-Herder-Str.\ 3,
D-79104 Freiburg, Germany}

\affiliation{EUCOR Centre for Quantum Science and Quantum Computing, Albert-Ludwigs-Universit\"at Freiburg, Hermann-Herder-Str.3, D-79104 Freiburg, Germany}

\author{Andrew Forbes}

\affiliation{School of Physics, University of the Witwatersrand, Private Bag 3, Johannesburg 2050, South Africa}

\begin{abstract}
\noindent
\hlstart
Structured light offers wider bandwidths and higher security for communication.
However, propagation through complex random media, such as the Earth's atmosphere, typically induces intermodal crosstalk.
We show numerically and experimentally that coupling of photonic orbital angular momentum (OAM) modes is governed by a \emph{universal} function of a single parameter -- the ratio between the random medium's and the beam's transverse correlation lengths, even in the regime of pronounced intensity fluctuations.

\hlend
\end{abstract}
\maketitle

\vspace{\columnsep}
\twocolumngrid
\noindent\emph{Introduction.---}
\label{sec:intro}
\hlstart
Optical communication strives to answer the growing demand of high-bandwidth links \cite{Lavery18, Puttnam21, Zhou21a}.
In particular, photonic spatial degrees of freedom such as \emph{orbital angular momentum} (OAM) offer an unbounded Hilbert space to encode information,
and provide intrinsic support for quantum key or entanglement distribution protocols \cite{Pors11, Leach12, Fickler12,Otte:20}, which with high-dimensional multiplexing enables greater channel capacities \cite{Wang12} and enhanced security \cite{zeilinger06}.
While communication based on such \emph{twisted photons} was successfully demonstrated for table-top \cite{mirhosseini2015}, indoor \cite{Vallone14} and short outdoor \cite{Sit:17} channels, transport of photonic OAM through \emph{complex random media}, e.g., the atmosphere \cite{Pors11a, Malik12, Ibrahim13}, water \cite{Ren:2016iv,Bouchard:18} or multimode fiber \cite{Zhou21a,popoff2021}, remains challenging:
stochastic fluctuations of the underlying media's refractive index induce phase distortions as well as intensity fluctuations upon propagation, leading to transmission losses and power transfer from information encoding modes to others --  \emph{intermodal crosstalk} -- which hinders reliable identification of the input modes \cite{Paterson05,Cox21}. Hence, for a successful optical communication in complex random media
we need to better
understand intermodal crosstalk therein.

To that end, we consider OAM-carrying Laguerre-Gaussian (LG) modes, which are most commonly used for classical and quantum communication
\cite{Lavery18,Zhou21a,Pors11, Leach12, Fickler12,Otte:20,mirhosseini2015,Vallone14,Sit:17,Pors11a, Malik12, Ibrahim13,Ren:2016iv,Krenn16,Lavery17,Bouchard:18,Zhou21a,popoff2021,Paterson05,Cox21}, as well as Bessel-Gaussian (BG) modes \cite{McLaren:12, Doster16}.
Information is typically encoded into the mode's OAM associated with the azimuthal index $\ell\in\mathbb{Z}$
\cite{Allen92};
in addition, the transverse amplitude distribution furnishes another degree of freedom characterized for LG \cite{Karimi2014,Zhou:17} and BG \cite{Gori87} modes, respectively, by the discrete radial index $p\in\mathbb{N}_0$ and the continuous radial wave number $\beta \in \mathbb{R}^+$.
 
Their rich transverse intensity structures bring about various applications of twisted photons. 
For example, LG modes with $p=0$ have the minimal space-bandwidth product and are usually employed in OAM-multiplexed systems \cite{Zhao:2015kx}. 
Furthermore, the doughnut-like intensity distribution of such modes is relatively robust under perturbations \cite{Krenn16,Lavery17}.
On the other hand, LG modes with $p>0$ allow for a substantial increase of the bandwidth capacity
\cite{Trichili:2016df}.
As for BG modes, they are promising due to their self-healing \cite{McLaren:2014ij} and resilience \cite{Doster16,Sorelli:19} properties. 

\noindent\onecolumngrid
    \hspace{-10pt}\begin{minipage}{\linewidth}
        \begin{figure}[H]
            \centering
            \includegraphics[width=1\linewidth]{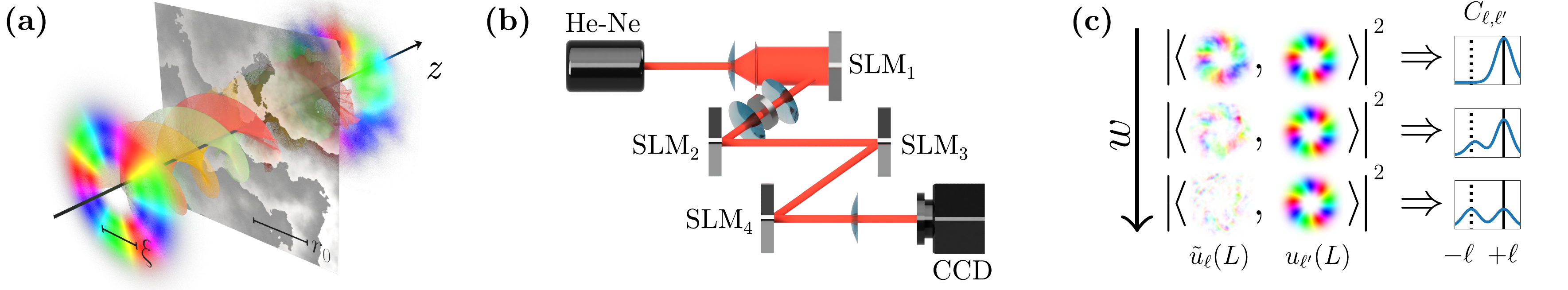}
            	\caption{
		(a) Incident LG$_{p=0}^{\ell=3}$ beam with
        transverse correlation length $\xi$ propagating through
        a random medium represented by a phase screen with correlation length $r_0$.
		 (b) Experimental implementation of the split-step method via spatial light modulators (SLMs). 
		(c) Qualitative illustration of crosstalk  $C_{\ell,\ell'}$ of an incident LG$_{p=0}^{\ell=3}$ mode, after propagation through a random medium to $z=L$, and of the vacuum-propagated LG$_{p=0}^{\ell'}(z=L)$ modes, for increasing distortion strength $w=w_0/r_0$ from top to bottom. Original and opposed OAM, i.e., $\pm\ell$, are indicated by the solid and dashed black lines, respectively.}
	\label{fig:lg}
        \end{figure}    
    \end{minipage}
\vspace{\columnsep}
\twocolumngrid
\noindent In this Letter, we show that twisted photons, i.e., LG and BG modes, exhibit a universal dependence of the crosstalk between modes with opposite OAM governed by the transverse length scale ratio of the beam and medium.

\emph{The model.---}
\label{sec:model}
We consider transverse optical modes $\psi({\bf r})$ arising within the paraxial approximation as solutions of the \emph{parabolic wave equation} describing free diffraction
\cite{Andrews05}. 
Twisted light modes are cylindrically symmetric solutions of this equation \cite{Forbes_2021}.
They feature distinct phase and intensity profiles which may be characterized by the transverse \emph{correlation length} $\xi$ \cite{Leonhard15} and by their $\ell$ times intertwined helical phase fronts along the propagation axis $z$, as illustrated in Fig.~\ref{fig:lg}(a).

However, photonic phase fronts are fragile under inherent refractive index fluctuations of the atmosphere or of other complex media.
Here, we consider the propagation of a monochromatic laser beam through clear atmospheric channels,
or through non-absorbing Gaussian media. In either case, the typical size of refractive index inhomogeneities is much larger than the laser wavelength, so that wave scattering is mainly in the forward direction \cite{Andrews05,Ishimaru78}. We choose the beam's wavelength 
to match the infrared transparency window of atmospheric turbulence \cite{Andrews05}, where wave attenuation due to absorption and multiple scattering by molecules and aerosols over propagation distances of a few kilometers is negligible \cite{Andrews05, meanfreepath}.
In other words, our propagation distance $L$ in the atmosphere is much shorter than the light's mean free path $l$ \cite{Carminati21, meanfreepath}, and the same regime is assumed for the Gaussian medium.
The beam's remaining sensitivity to refractive index fluctuations is modeled by the \emph{stochastic} parabolic equation
which incorporates random noise induced by the medium \cite{Andrews05,Segev:2013uq}.
To find its solution, we employ the \emph{split-step method} \cite{Schmidt10, Chatterjee14}, which relies on segmenting the entire propagation path into discrete, medium-induced phase modulations, i.e., \emph{phase screens}, see Fig.~\ref{fig:lg}(a), interconnected by free diffraction
\cite{Born99,Goodman05,Schmidt10}.

\begin{figure}
	\includegraphics[width=\columnwidth]{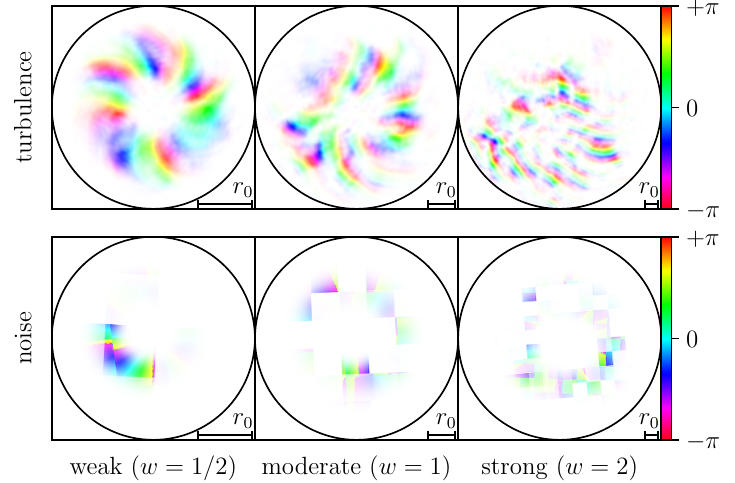}
	\caption{ 
	Intensity-weighted transverse phase profiles of an incident LG$_{p=0}^{\ell=5}$ mode, after propagation through Kolmogorov turbulence (top), or through Gaussian noise (bottom), for three different distortion strengths $w=1/2, 1, 2$, corresponding to $\sigma_R^2=0.24, 0.76, 2.42$ from left to right.}
	\label{fig:modes}
\end{figure}

The phase screens
incorporate
the medium's transverse correlation length.
For atmospheric \emph{Kolmogorov turbulence} this characteristic length scale is known as the \emph{Fried parameter} $r_0$ \cite{Andrews05}; the ratio $w:=w_0/r_0$, with the \emph{beam waist} $w_0$, quantifies the resulting \emph{distortion strength}.
Similarly, the \emph{Rytov variance} $\sigma_R^2$ quantifies intensity fluctuations and distinguishes between \emph{weak scintillation} $\sigma_R^2<1$
and \emph{strong scintillation} $\sigma_R^2\geq1$ \cite{Ishimaru78}.
We ensure that every propagation segment satisfies $\sigma_R^2<1$.
Furthermore, we represent more general random media by \emph{Gaussian noise} based on normally distributed block matrices with block size $r_0$, independently altering both amplitude and phase of given transverse modes, where the block size $r_0$ mimics the correlation length of Kolmogorov turbulence.
In this case, we suppress systematic anisotropies, e.g., due to distinct lengths of a block's side and diagonal, by randomly shifting and rotating the resulting noise grid within the transverse plane.
Examples of a perturbed LG beam are shown in Fig.~\ref{fig:modes}, where we distinguish three regimes of distortion: weak, moderate, and strong, quantified by $w < 1$, $w\approx1$, and $w < 1$, respectively.

\begin{figure*}[ht]
\centering
\includegraphics[width=\linewidth]{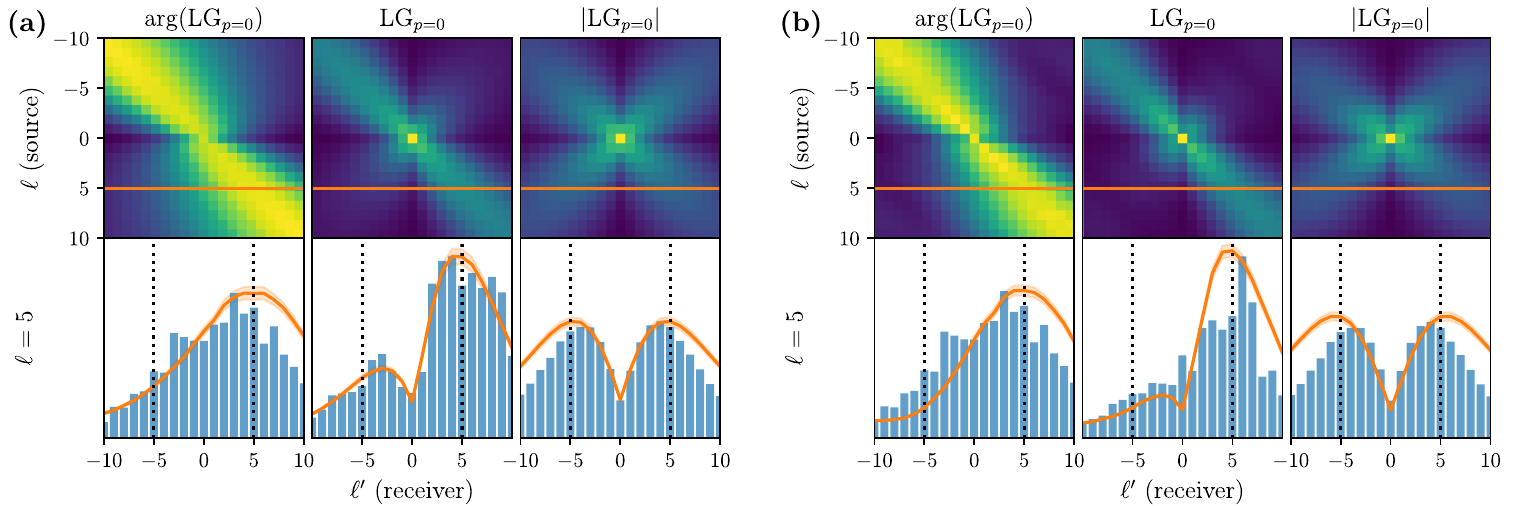}
\caption{Numerically obtained average (2500 realizations) crosstalk matrices in moderate $(w=1)$ Kolmogorov turbulence (a) and Gaussian noise (b).
The crosstalk matrices (top) are obtained when projecting
the image $\tilde{u}_\ell(L)$ of the incident mode $\text{LG}_{p=0}^\ell$, onto vacuum-propagated modes $u_{\ell'}(L)=\text{LG}_{p=0}^{\ell'}(z=L)$ in the middle columns, according to Eq.~\eqref{eq:ct}. Left and right columns represent the beam's phase and intensity profiles' modifications
quantified by substitution of arg$[$LG$_{p=0}^{\ell'}(z=L)]$ and $|$LG$_{p=0}^{\ell'}|$ for $u_{\ell'}(L)$ in Eq.~\eqref{eq:ct}, respectively.
Rows $C_{\ell=5,\ell'}$ (bottom) depict numerical (orange curve, error bands give one standard deviation) as well as experimental results (blue bars, 60 realizations). The crosstalk matrices are normalized such that $\sum_{\ell,\ell'=-10}^{10}C_{\ell,\ell'}=1$.}
\label{fig:ct}
\end{figure*}

To benchmark our theoretical description, we experimentally realize a turbulent link by downscaling realistic channels to tabletop dimensions \cite{Rodenburg14}. The setup, shown in Fig.~\ref{fig:lg}(b), allows to experimentally simulate propagation through weak to strong scintillation.  The salient elements are divided into three stages.  In the generation stage, a He-Ne laser beam
is expanded and collimated before being directed onto a reflective
\emph{spatial light modulator} (SLM$_1$) which generates the desired source mode. This mode then enters the distorting section of the setup where it passes through a random medium implemented by a two stage split-step propagation, with phase shifts programmed to SLM$_2$ and SLM$_3$, each followed by 1\,m of free diffraction.  This 2\,m laboratory system maps to a $L=200$\,m real-world channel with Rytov variances of up to $\sigma_R^2 \approx 1.8$, matching the numerically investigated conditions.
In the final stage, the modal decomposition is performed optically \cite{pinnell2020modal} with the aid of a match filter programmed to SLM$_4$ and a Fourier transforming lens, with the on-axis intensity measured with a camera (CCD detector).

\emph{Crosstalk among distorted modes.---}
\label{sec:crosstalk}
The probability of identifying a source mode's original OAM $\ell$ after propagation as $\ell'$ is quantified via the \emph{crosstalk matrix}
\begin{equation}
	C_{\ell,\ell'}(L):=|\langle \tilde{u}_\ell(L), u_{\ell'}(L) \rangle|^2,
	\label{eq:ct}
\end{equation}
where $\tilde{u}_{\ell}(L)$ and $u_{\ell'}(L)$ are complex modes \footnote{The full mode functions read $u_{p=0, \ell}(\rho, \varphi, z=L)$, where the radial indices $p=0$ and the transverse coordinates $\rho, \varphi$ were suppressed for brevity.} propagated across a random medium and through a vacuum channel each of length $L$, respectively, and $\langle \cdot, \cdot \rangle$ denotes the standard scalar product in the transverse space at $z=L$ \cite{supp}.
The distorted modes $\tilde{u}_\ell(L)$ are connected to the incident modes $u_\ell$ by the transmission operator $T(L)$, which approximates the medium's scattering matrix \cite{Beenakker97a} whenever wave reflection is neglected. Although geometric truncation due to finite size apertures induces non-unitarity of  $T(L)$, its eigenvalues in weak turbulence exhibit the bimodal distribution \cite{Bachmann2023} that characterizes complex scattering media \cite{Beenakker97a}. Accordingly, the crosstalk matrix can be interpreted as the square modulus of the elements of the transmission operator's matrix representation in the bases of the incident modes and their vacuum-propagated images.

Consequently, higher transmission fidelities correspond to weaker crosstalk, i.e., to matrices $C_{\ell,\ell'}(L)$ with prevailing diagonals.
Figure~\ref{fig:lg}(c) sketches the crosstalk amplitudes for a fixed source mode
LG$_{p=0}^{\ell=3}$ after propagation through a channel of length $L$, for three distortion regimes quantified by $w$.
As observed previously \cite{Klug21}, under weak distortion (top row)
the
detection probability of the original OAM $\ell=3$ (solid black line)
is highest and crosstalk is limited to neighboring modes.
In contrast, under strong distortion, see bottom row of Fig.~\ref{fig:lg}(c), we find a $\pm\ell$-symmetric distribution of the crosstalk amplitudes with separate maxima close to $\pm\ell$ (black lines).
In this case, the mode's phase front is destroyed, while its amplitude -- which is on average independent of the sign of OAM -- partially survives.
Finally, the moderate distortion regime, see center row of Fig.~\ref{fig:lg}(c), is of particular interest: It represents the transition between rather faithful OAM transmission (top row)
and dominant phase destruction (bottowm row).
Both, the beam's amplitude and phase profiles
are altered, but neither is completely destroyed.

To elucidate the origin of the disorder-induced crosstalk between OAM modes, we decompose the incident LG$_{p}^{\ell}$, yielding $\tilde{u}_\ell(L)$ in Eq.~\eqref{eq:ct} after transmission through moderate, $w=1$, disorder at $z=L$, into the vacuum-propagated LG modes by setting $u_{\ell'}(L)=\text{LG}_{p}^{\ell'}(z=L)$ in Eq.~\eqref{eq:ct}
as illustrated for $p=0$ in Fig.~\ref{fig:ct}.
Furthermore, we quantify the channel's imprint on the transmitted beam's phase and intensity profile alone by
by setting $u_{\ell'}(L)$ to
arg$[\text{LG}_p^{\ell'}(z=L)]$ and $|\text{LG}_p^{\ell'}(z=L)|$ in Eq.~\eqref{eq:ct}, respectively, see left and right columns of Fig.~\ref{fig:ct}(a,b).
Note that, in general, random media also induce coupling to modes with $p' \neq p$, but the prescribed projection onto individual radial indices is common in communication scenarios \cite{Bouchard:18a,Sorelli19}, and required when finite-size apertures geometrically truncate modes with larger $p$, or if both, $p$ and $\ell$, are used for information encoding.

\begin{figure*}[ht]
\centering
\includegraphics[width=\linewidth]{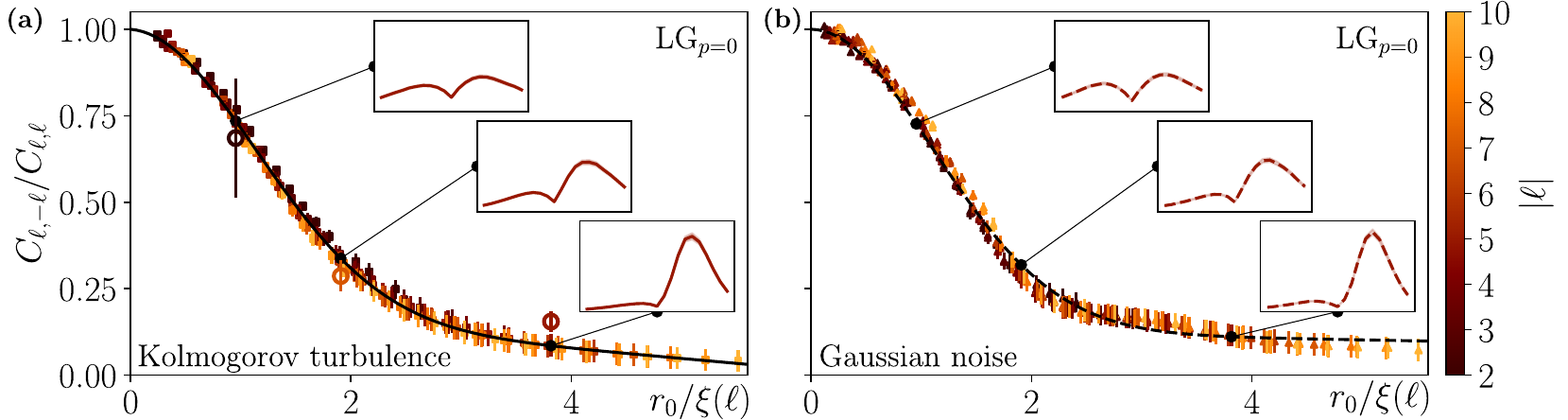}
\caption{
Ratio of average (2500 realizations) crosstalk between two OAM-opposed (i.e. $\ell$ and $-\ell$) modes propagated through numerically simulated Kolmogorov turbulence (a) and Gaussian noise (b) versus the ratio of the medium's and beam's transverse coherence lengths for a range of $|l|$.
The incident LG$_{p=0}^{\ell}$ were projected onto LG$_{p=0}^{\mp\ell}(z=L)$ at the receiver side. The emerging universal curve is fitted with a Gaussian (black curves) with parameters in \cite{fitparametersol}. Circles with error bars in (a) represent experimentally measured average (60 realizations) crosstalk in Kolmogorov turbulence.
Insets
illustrate rows at $\ell = 5$ of
corresponding crosstalk matrices (same axes as in
Fig.~\ref{fig:ct}). Error bands or bars
give one standard deviation.
}
\label{fig:uni}
\end{figure*}

The phase profile, see top left plots in Fig.~\ref{fig:ct}(a,b), of the crosstalk matrix
concentrates about the diagonal, which manifests
the prevailing coherent coupling of modes with matching OAM, i.e., $\ell'=\ell$, confirming previous findings \cite{Klug21}. Furthermore, the diagonal's broadening attests to disorder-induced coupling among modes with different OAM, i.e., $\ell' \neq \ell$.
In contrast, when projecting onto the amplitude profiles, see top right plots in Fig.~\ref{fig:ct}(a,b), crosstalk matrices exhibit pronounced, symmetric diagonal and \emph{anti-diagonal} structures. Their symmetry is expected because the amplitude of LG modes is independent of the sign of OAM, and their shape is very suggestive 
when considering the projection onto LG modes, see top middle plots in Fig.~\ref{fig:ct}(a,b).
Coherent coupling of neighboring modes, due to matching OAM phase is combined with the anti-diagonal crosstalk originating from amplitude overlap, leading to the presence of damped anti-diagonal elements in agreement with Ref.~\cite{Sorelli19}.

The lower row of Fig.~\ref{fig:ct} compares experimental measurements (blue bars)
with numerical results (orange curves)
in form of crosstalk matrix rows which describe the detection probabilities of various $\ell'$ when transmitting $\ell=5$, cf. orange lines in the upper plots.
Experimental results qualitatively agree with numerical results, and their systematic underestimation for larger $\ell'$
originates from the receiving aperture
damping the overall transmission of these correspondingly wider modes \cite{Andrews05}.
In the bottom middle plots of Fig.~\ref{fig:ct}(a,b), i.e., when projecting onto full LG modes, 
we observe an \emph{asymmetric doublet} whose second diminished peak is systematically shifted toward the center.
While crosstalk between $\pm\ell$ is still amplified, this behavior further highlights that we do not observe direct $\pm\ell$-coupling of LG modes but rather a tradeoff between destruction of OAM-preserving 
phase and surviving symmetric amplitude.

We note that \emph{anti-diagonal}, i.e., $\pm\ell$, crosstalk becomes weaker for larger $|\ell|$, cf. fading of anti-diagonals in the top middle plots of Fig.~\ref{fig:ct}(a,b).
This observation is consistent with well-known results that entangled photonic OAM qubit states with opposite $\ell$ become more robust in turbulence as $|\ell|$ increases \cite{Smith06}.
Careful study of LG modes \cite{Leonhard15} within the weak scintillation regime attributed this enhanced robustness to the finer transverse spatial structure of such photons characterized by their analytically given phase correlation length $\xi$.
The latter gives the average distance between two points in the transverse profile of LG beams with a phase difference of $\pi/2$  \cite{Leonhard15, Yang21}.

Due to its ubiquity in communication protocols \cite{Ndagano17,Vallone14,Ibrahim13,Leonhard15,Smith06,Bachmann19}, we further explore
the anti-diagonal crosstalk for a range of distortion strengths $w$ and azimuthal indices $\ell$ of incident OAM modes.
To this end, we investigate the ratio between anti-diagonal and diagonal crosstalk, i.e., $C_{\ell,-\ell}$ over $C_{\ell,\ell}$.
For LG modes, this ratio is plotted in Fig.~\ref{fig:uni}, versus the medium's transverse correlation length $r_0$ normalized by 
$\xi$ for
Kolmogorov turbulence
and Gaussian noise.
Further modes are considered in the Supplemental Material \cite{supp}.
Remarkably, the rescaling of $r_0$ collapses the data onto a \emph{universal}, i.e., $\ell$-independent, crosstalk curve -- even in the regime of strong scintillation.
Given the analogy between optical wave transmission and electronic transport
\cite{akkermansbook}, this  behavior is reminiscent of the \emph{universal conductance fluctuations}
observed
for electrons
in solid state physics \cite{Beenakker97a},
where scattering properties are a universal function of $L/l$. 
However, the latter universality occurs in the diffusive regime, $L/l\gg 1$, which is opposite to the one here considered.

To ease the comparison and to highlight the universality, the data points were fitted by a Gaussian \cite{fitparametersol}
with decaying offset, to mimic the slow systematic noise reduction with decreasing distortion strength.
It is notable that the two considered, i.e., Kolmogorov and Gaussian, media -- albeit fundamentally different -- result in a very similar crosstalk decay,
which is attributed to their similar transverse length scale.

The crosstalk ratio in Fig.~\ref{fig:uni} reflects the crossover from symmetric doublets
in strong distortion, where OAM-encoding phase information is destroyed, to dominant direct $\ell$-coupling in weak distortion.
Moreover, the numerical results are in quantitative agreement with experimental data for Kolmogorov turbulence, see circles in Fig.~\ref{fig:uni}(a), in the moderate and strong turbulence regime; the measurement for weak turbulence, see rightmost circle,
shows a systematic offset due to the decreased signal-to-noise ratio of the diminished anti-diagonal crosstalk in this case.
The universality of crosstalk for both Kolmogorov turbulence and generic Gaussian noise suggests that
such agreement can also be expected for
other random media.

\emph{Discussion.---}
The universal dependence of crosstalk amplitudes between modes of opposite OAM on the rescaled Fried parameter was previously predicted in the regime of weak scintillation,
for Kolmogorov \cite{Leonhard15} and non-Kolmogorov turbulence \cite{Bachmann19} alike, as well as under deterministic perturbation of twisted photons by angular apertures \cite{Sorelli20a}. Moreover, it was established that in this
regime bipartite entanglement of
photonic qubit states with opposite OAM depends on the same parameters as crosstalk -- and, hence, exhibits universal behavior as well \cite{Leonhard15,Bachmann19,Sorelli20a}. However, entanglement of twisted photons depends on intensity fluctuations \cite{Roux15} and this
sensitivity renders rescaling into a universal entanglement evolution impossible under strong scintillation. 
Our present theoretical and experimental results show that \emph{crosstalk} among structured light modes is nonetheless universal beyond weak scintillation conditions.
Importantly, only the characteristic transverse correlation length of the medium-induced errors needs be known to understand the impact of modal scattering on twisted light modes, with clear impact in imaging and communication through noisy channels.
This result in particular implies
that the ubiquitous superposition 
of $\pm\ell$
\cite{Ndagano17,Vallone14,Ibrahim13,Leonhard15,Smith06,Bachmann19}, is
far from an optimal choice of encoding information.

\emph{Conclusion.---}
In this Letter, we have studied the crosstalk of twisted photons in Kolmogorov turbulence and Gaussian noise. In both cases, we have numerically as well as experimentally confirmed pronounced crosstalk among LG modes of opposed OAM. Instead of originating from direct OAM cross coupling, this behavior was identified as a tradeoff effect between matching intensity patterns and destroyed phase information.
Moreover, we have uncovered a universal crosstalk decay complying with experimental measurements by setting the random media's transverse correlation length into relation with the beam's phase structure.
We have established that the universality holds for LG modes with different radial indices, that is, with different transverse amplitude distributions, as well as for BG modes with different beam waists \cite{supp}. We therefore envisage a similar behavior for other sets of OAM modes upon a suitable generalization of the phase correlation length \cite{Leonhard15,supp}.

Our results
may lead to novel venues 
for communication.
In particular,
optimizing the beam's transverse correlation length for given distortion strengths will diminish the crosstalk, irrespective of the precise nature of the underlying random medium itself. As we have demonstrated, this can be achieved by suitable choice of $\ell$ or $p$  ($\beta$) for LG (BG), or, due to the scaling properties of LG and BG modes, by adapting the beam waist $w_0$.

\emph{Acknowledgements.---}
D.B. acknowledges financial support by the Studienstiftung des deutschen Volkes.
The authors acknowledge support by the state of Baden-Württemberg through bwHPC
and the German Research Foundation (DFG) through grant no INST 40/575-1 FUGG (JUSTUS 2 cluster).
G.S. acknowledges partial funding by French ANR (ANR-19-ASTR0020-01). V.S. and A.B. acknowledge partial funding and support through the Strategiefonds der Albert-Ludwigs-Universität Freiburg and the Georg H. Endress Stiftung.

\bibliography{singularmodes}

\begin{thebibliography}{61}
\bibitem [{mea()}]{meanfreepath}%
  \BibitemOpen
  \href@noop {} {\bibinfo  {journal} {At a wavelength of $\lambda=1550\,$nm,
  about 99\% of the incident power is transmitted over a distance of $1\,$km
  \cite{Andrews05}. From the Lambert-Beer law, this corresponds to the mean
  free path $l \simeq 100\,$km \cite{Carminati21,Beenakker97a}}\ }\BibitemShut
  {NoStop}%
\end{thebibliography}

\clearpage
\section{Supplemental Material}
\noindent In this Supplemental Material,
we present the universal crosstalk ratio for \textit{Laguerre-Gaussian} (LG) modes with radial index $p\geq0$ as well as \textit{Bessel-Gaussian} (BG) modes.
Furthermore, we elaborate on the definition of the crosstalk matrix and provide details for the phase correlation length that is vital for rescaling.

\emph{Definition of crosstalk matrices.---}
\label{sec:crosstalk}
The crosstalk matrix, as defined in Eq.~(1) of the main manuscript, allows to
quantify the overlap between two sets of transverse modes -- typically,
perturbed and unperturbed (reference) modes.  
First, we denote by $\tilde{u}_\ell(\rho, \varphi, z=L)$ a mode set 
that is initially (i.e., 
at propagation distance $z=0$)
prepared with indices $\ell$ and 
propagated to $z=L$ through a random medium. 
Second, we denote by 
$u_{\ell'}(\rho, \varphi, z=L)$ 
a set of 
reference modes that is initially prepared with 
indices $\ell'$ and subsequently propagated 
to $z=L$ in vacuum. 
Finally, the crosstalk matrix elements are defined as the modulus squared
of the standard scalar product of these two mode sets in the transverse plane at $z=L$, that is
    \begin{equation}
        C_{\ell, \ell'}(L) = \Bigg|\int_{0}^{R} \hspace{-5pt} \rho  d \rho \int_{0}^{2\pi}\hspace{-5pt}d \varphi\,\, \tilde{u}^*_\ell(\rho, \varphi, L)\hspace{-2pt}\, u_{\ell'}(\rho, \varphi, L) \Bigg|^2,
        \label{eq:ct2}
    \end{equation}
where $\rho, \varphi$ denote the usual polar coordinates and $R$ is the radius of the receiver aperture at $z=L$.

\begin{figure}[t!]
	\includegraphics[width=\columnwidth]{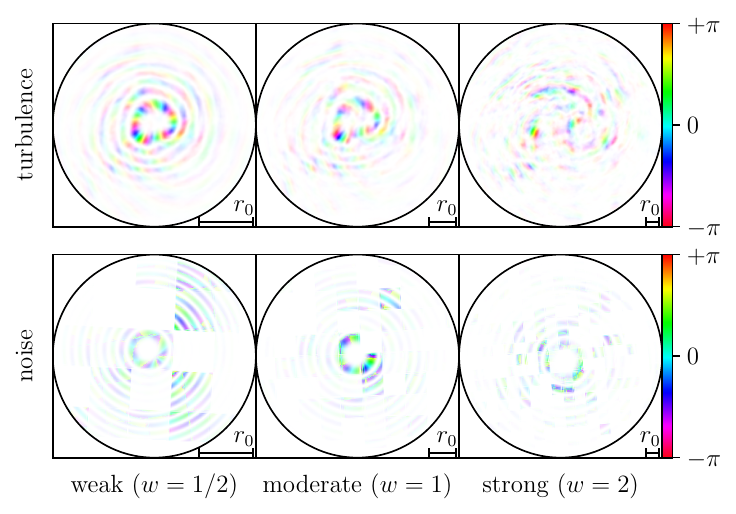}
	\caption{ 
	Intensity-weighted transverse phase profiles of an incident BG$_{\beta}^{\ell=5}(z=0)$ mode with a beam waist of $5w_0$ (i.e., five times the beam waist of the previously considered LG modes) after propagation through Kolmogorov turbulence to $z=L$ (top) or Gaussian noise (bottom), for three different distortion strengths $w=1/2, 1, 2$, corresponding to $\sigma_\text{R}^2=0.24, 0.76, 2.42$ (left to right).}
	\label{fig:bgmodes}
\end{figure}

\emph{Considered OAM modes.---}
In our case, the index $\ell\in\mathbb{Z}$ describes OAM and we consider LG and BG modes as a prime examples of twisted light \cite{Allen92}. Accordingly, 
$u_\ell(\rho, \varphi, z) = \text{LG}_{p}^{\ell}(z)$ or 
$\text{BG}_{\beta}^{\ell}(z)$ in Eq.~\eqref{eq:ct2}.
In addition to OAM, LG and BG modes
possess radial structure characterized by the discrete index
$p\in\mathbb{N}_0$ and the transverse wave number $\beta\in \mathbb{R}^+$, respectively. 
The integer
$p+1$ gives the number of concentric intensity rings (for $\ell=0$ including a dot in the center) in the transverse plane of an LG beam;  
the continuous parameter $\beta$ corresponds to the spatial frequency of concentric rings in the transverse plane of a BG beam (cf. Fig.~\ref{fig:bgmodes}).
The modes' explicit form at $z=0$ with beam waist $w_0$ is given by \cite{Andrews05, Gori87}
\begin{align}
    \text{LG}_p^\ell(\rho, \varphi, z=0) &= \frac{1}{\sqrt{2\pi}}\, R_p^{|\ell|}(\rho)\, \exp(i\ell \varphi),\nonumber\\
    \text{BG}_\beta^\ell(\rho, \varphi, z=0) &= \frac{1}{\sqrt{2\pi}}\, S_\beta^{\ell}(\rho)\, \exp(i\ell \varphi),
    \label{eq:modes}
\end{align}
with their respective radial functions
\begin{align}
    R_p^{|\ell|}(\rho) &= C_p^{|\ell|}\left(\frac{\sqrt{2}\rho}{w_0}\right)^{|\ell|}L_p^{|\ell|}\left(\frac{2\rho^2}{w_0^2}\right)\exp\left(-\frac{\rho^2}{w_0^2}\right),\nonumber\\
    S_\beta^{\ell}(\rho) &= D_\beta^{|\ell|}\, J_{\ell}(\beta \rho)\,\exp\left(-\frac{\rho^2}{w_0^2}\right),
    \label{eq:radialparts}
\end{align}
where $C_p^{|\ell|}$ and $D_\beta^{|\ell|}$ are normalization constants
\footnote{Normalization constants for LG and BG modes are given by $C_p^{|\ell|}  = \frac{2}{w_0}\,\sqrt{\frac{p!}{(p+|\ell|)!}}$ and $D_\beta^{|\ell|}  = \frac{2}{w_0}\, \frac{\exp(w_0^2\beta^2/8)}{\sqrt{I_{|\ell|}(w_0^2\beta^2/4)}}$, respectively, where $I_\ell(\cdot)$ denotes the modified Bessel function of the first kind.}, while $L_p^{|\ell|}(\cdot)$ and $J_\ell(\cdot)$ denote the associated Laguerre polynomials and the Bessel function of the first kind, respectively.
Both mode sets are normalized, that is, their total intensity given by the modulus squared of the mode functions provided in Eq.~\eqref{eq:modes}, yields unity after integration over the entire transverse plane. Furthermore, we decompose these modes into their amplitude and phase parts.
We then consider separately their phase, i.e., $\arg(\text{LG}_{p}^{\ell})$ or $\arg(\text{BG}_{\beta}^{\ell})$,
and for LG modes also their amplitude, $|\text{LG}_{p}^{\ell}|$, 
see, e.g., Fig.~3 of the main manuscript.

\emph{Definition of the phase correlation length.---}
The phase correlation length $\xi$ of an OAM beam
is the characteristic distance between two points in the transverse plane with a phase difference of $\pi/2$ weighted by the mode's intensity profile \cite{Leonhard15}. This quantity allows us to uncover the universal nature of the crosstalk in random media when $\xi$ is compared with the medium's transverse correlation length.

This length scale was introduced in Ref. \cite{Leonhard15} for LG modes with $p=0$ as
\begin{align}
    \xi(p,\ell) = \sin\left(\frac{\pi}{2|\ell|}\right)\, \int_0^\infty d\rho\, \rho^2 \,|R_{p}^{|\ell|}(\rho)|^2.
    \label{eq:xi}
\end{align}
Below, we apply Eq.~\eqref{eq:xi} to compute $\xi(p,\ell)$ for LG modes with $p\geq0$, as well as for BG modes [in which case we substitute $p$ by $\beta$ and $R_{p}^{|\ell|}$ by $S_{\beta}^{|\ell|}$, cf. Eq.~\eqref{eq:radialparts}]. We note that Eq.~\eqref{eq:xi} systematically overestimates the true phase correlation length in this case, because this expression neglects the phase flips by $\pi$ along a radial line in the transverse phase profile of LG or BG modes. Nevertheless, this effect is benign for moderate $p\lesssim 5$ (LG modes) or small frequencies such that $w_0\beta \lesssim 50$ (BG modes), which we use here to illustrate the universality of crosstalk, so that we can keep the original definition \eqref{eq:xi} of the phase correlation length. Furthermore, 
when the source modes are geometrically truncated by the finite-size input aperture, the upper integration limit in Eq.~\eqref{eq:xi} is set to the corresponding aperture radius $R_\text{src}$
and the final quantity has to be normalized by $\int_{0}^{R_\text{src}} \rho d\rho |R_{p}^{|\ell|}(\rho)|^2$.

\begin{figure*}[ht]
\centering
\includegraphics[width=\linewidth]{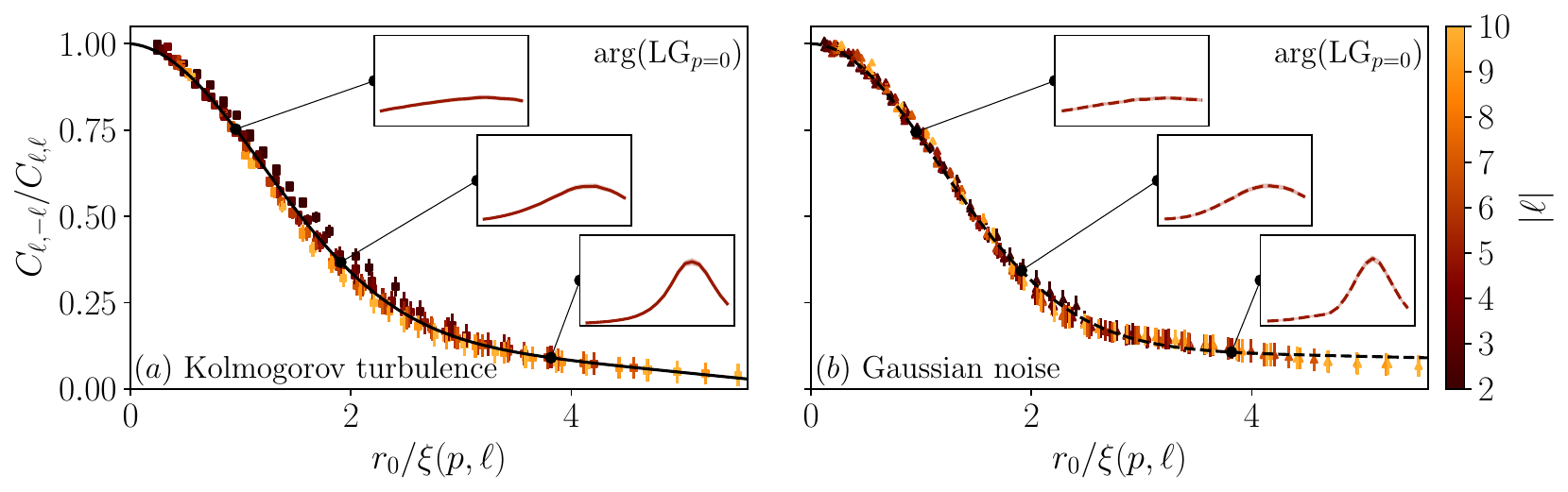}
\caption{Ratio of average (2500 realizations) crosstalk between two OAM-opposed (i.e. $-\ell$ and $\ell$) modes, propagated through numerically simulated Kolmogorov turbulence $(a)$ and Gaussian noise $(b)$, versus the ratio of the medium's and beam's transverse coherence lengths for a range of $|l|$ (see color bar legend). The incident LG$_{p=0}^{\ell}(z=0)$ were projected onto arg$[$LG$_{p=0}^{\mp\ell}(z=L)]$ at the receiver side. The emerging universal curve is fitted by a Gaussian decay [black curves, cf. Eq.~\eqref{eq:fit}] with parameters given in Tab.~\ref{tab:paras}. The channel geometry is the same as described in the model section of the main manuscript. Insets illustrate rows (at $\ell = 5$) of
corresponding crosstalk matrices (same axes as in the bottom
rows of Fig. 3 in the main manuscript). Error bands in the insets or bars in the main
plot correspond to one standard deviation.
}
\label{fig:p0-oam}
\end{figure*}

\emph{Universality for OAM with uniform intensity profile.---}
Figure~4 of the main manuscript demonstrates the universality of the ratio of OAM-opposed crosstalk $C_{\ell,-\ell}/C_{\ell, \ell}$ versus the rescaled Fried parameter $r_0/\xi(p, \ell)$ for $p=0$ LG modes, in both Kolmogorov turbulence and Gaussian noise. In Fig.~\ref{fig:p0-oam}, we show that this rescaling also leads to universal crosstalk when projecting onto arg[LG$_{p=0}^{\mp\ell}(z=L)$], i.e., OAM modes with uniform intensity profile. 
The observed universality is anticipated since the intensity profiles of LG modes are symmetric with respect to OAM reversal, such that the radial function $R_{p}^{|\mp\ell|}$ effectively becomes immaterial. This is also consistent with the result shown in Fig.~3 of the main manuscript where we have presented the crosstalk profiles.
For comparison and to highlight the universality, the crosstalk ratio was fitted in the same manner as described in the main manuscript, i.e., by a Gaussian with a decaying offset that mimics the slow decrease of systematic noise with decreasing turbulence strength:
\begin{equation}
    f(x)=(1-a)e^{-x^2/(2b^2)} + a - cx,
    \label{eq:fit}
\end{equation}
where $a,\;b,\;c\in \mathbb{R}$ are free fitting parameters which are summarized in Tab.~\ref{tab:paras}.

\begin{figure*}[ht]
\centering
\includegraphics[width=\linewidth]{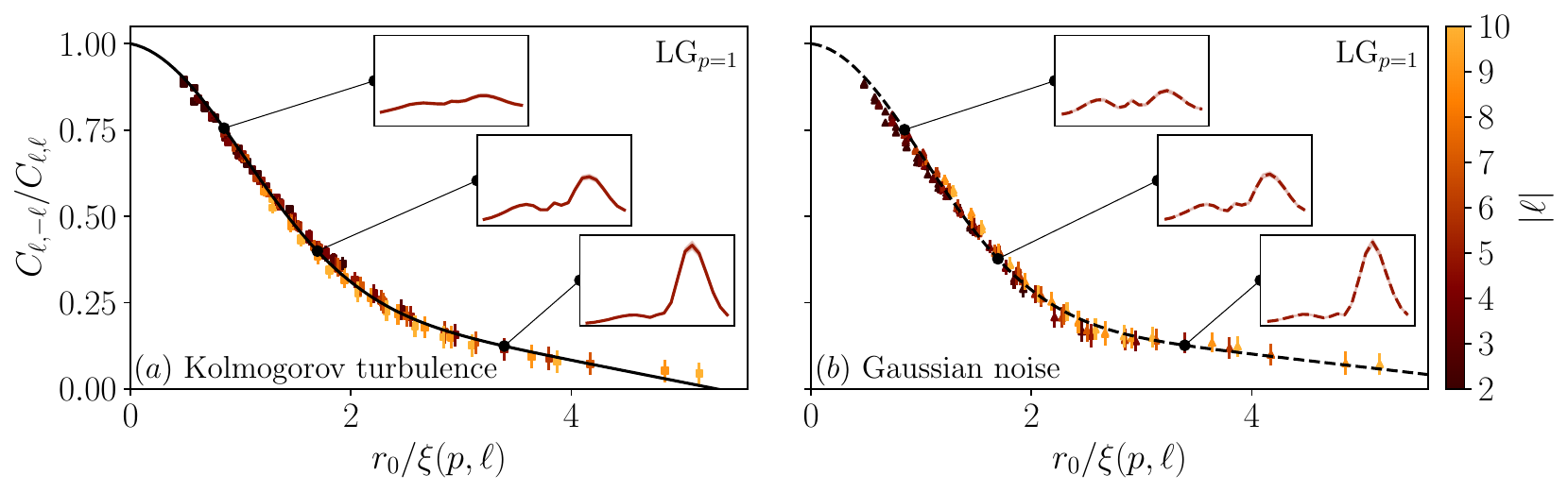}
\caption{Same as in Fig.~\ref{fig:p0-oam}, but for the incident modes LG$_{p=1}^{\ell}(z=0)$ projected onto LG$_{p=1}^{\mp\ell}(z=L)$ at the receiver side.} 
\label{fig:p1-lg}
\end{figure*}

\begin{figure*}[ht]
\centering
\includegraphics[width=\linewidth]{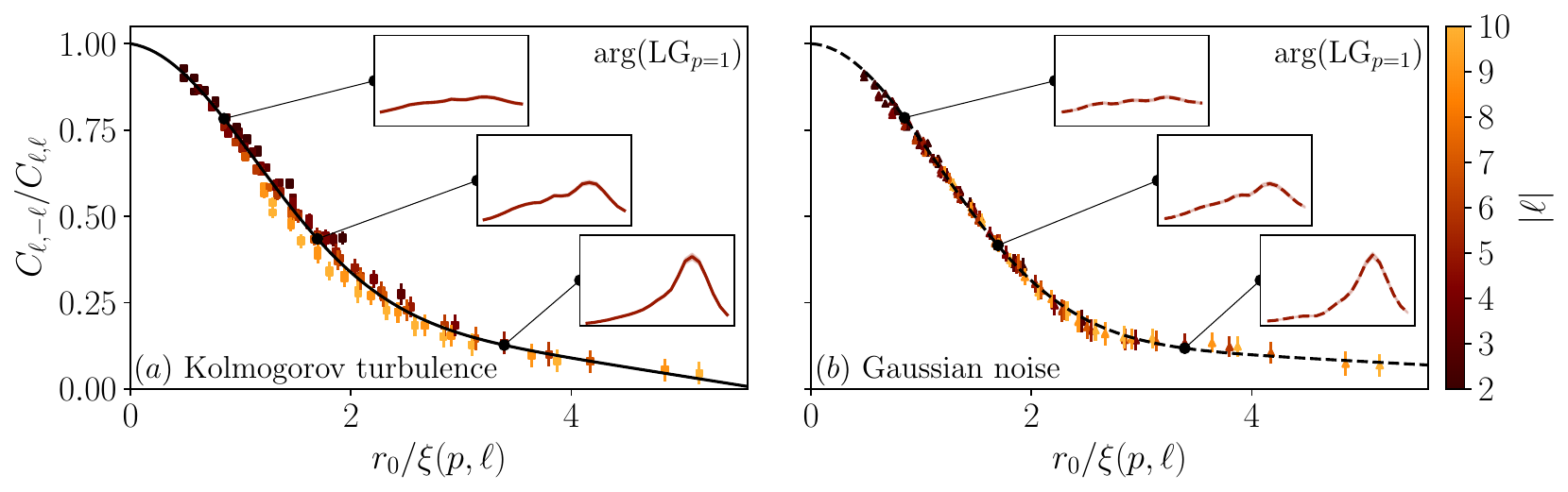}
\caption{Same as in Fig.~\ref{fig:p0-oam}, but for the incident LG$_{p=1}^{\ell}(z=0)$ projected onto arg$[$LG$_{p=1}^{\mp\ell}(z=L)]$ at the receiver side.}
\label{fig:p1-oam}
\end{figure*}

\begin{figure*}[ht]
\centering
\includegraphics[width=\linewidth]{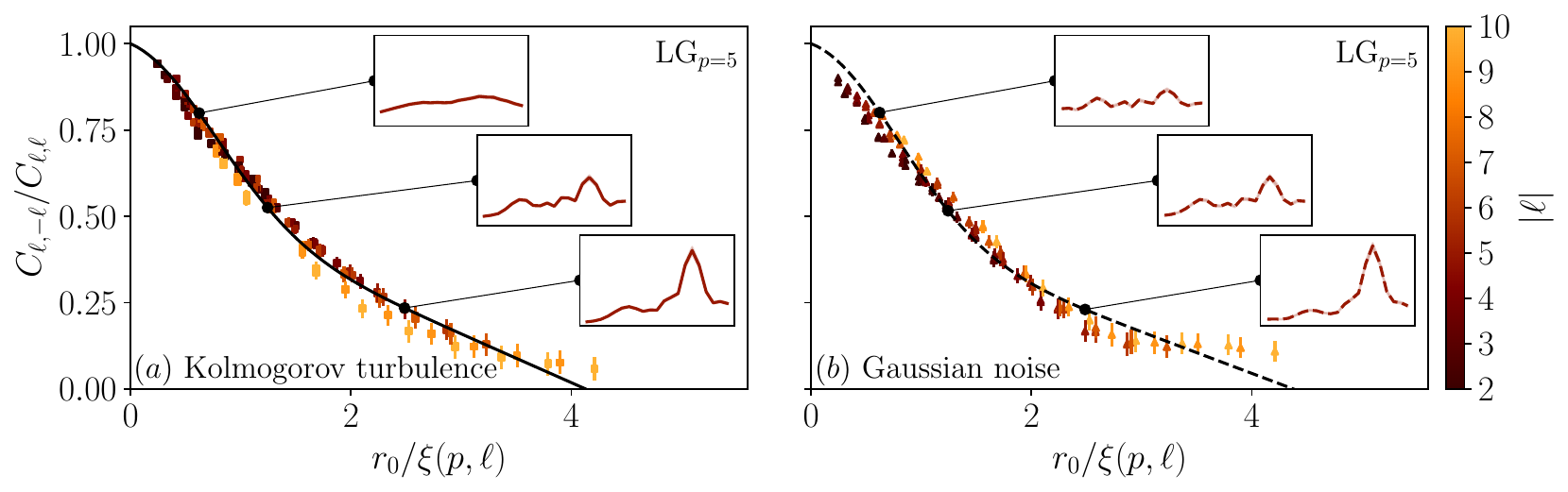}
\caption{Same as in Fig.~\ref{fig:p0-oam}, but for the incident LG$_{p=5}^{\ell}(z=0)$ projected onto LG$_{p=5}^{\mp\ell}(z=L)$ at the receiver side. The channel geometry is the same as described in the model section of the main manuscript but the aperture radius was increased by 30\,\% in order to accommodate the wider $p=5$ modes.}
\label{fig:p5-lg}
\end{figure*}

\begin{figure*}[ht]
\centering
\includegraphics[width=\linewidth]{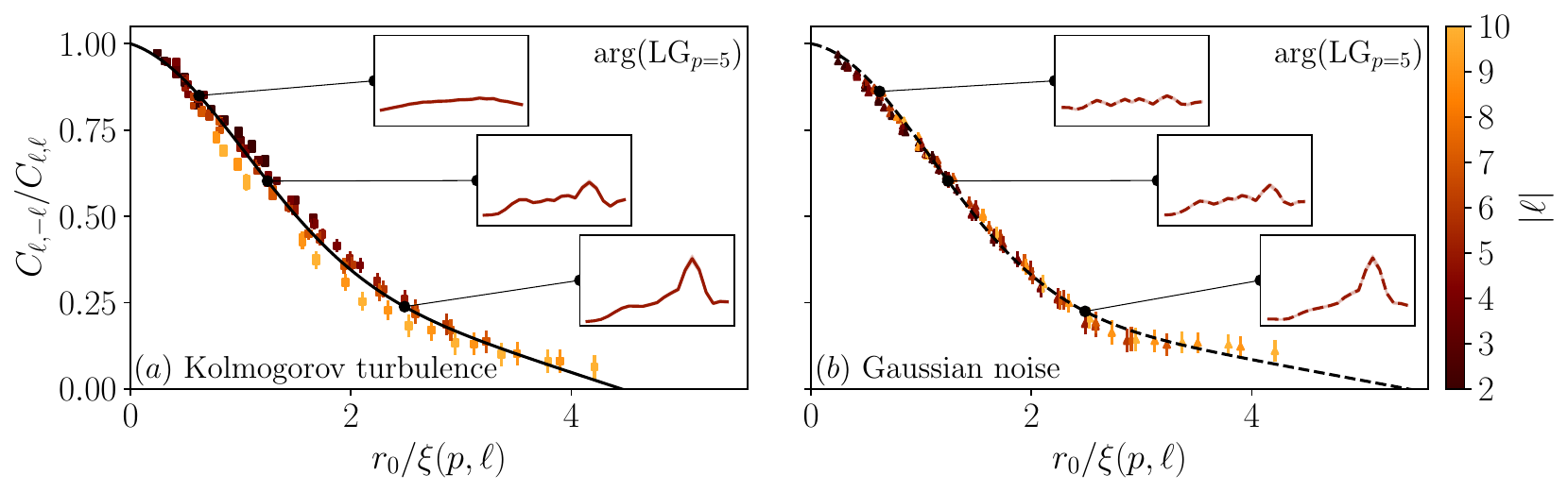}
\caption{Same as in Fig.~\ref{fig:p1-lg}, but for the incident LG$_{p=5}^{\ell}(z=0)$ were projected onto arg$[$LG$_{p=5}^{\mp\ell}(z=L)]$ at the receiver side. The channel geometry is the same as described in the model section of the main manuscript but the aperture radius was increased by 30\,\% in order to accommodate the wider $p=5$ modes.}
\label{fig:p5-oam}
\end{figure*}

\emph{Universality for higher order LG modes.---}
In the same way as for $p=0$, we consider LG modes with $p>0$. The crosstalk ratio when projecting onto the LG modes (cf. Fig.~4 of the main manuscript) is given for $p=1$ and $p=5$ in Figs.~\ref{fig:p1-lg} and \ref{fig:p5-lg}, respectively. The crosstalk ratio after projection onto the phase parts, i.e., arg$[\text{LG}_{p}^{\mp\ell}(z=L)]$, for $p=1$ and $p=5$ is presented in Figs.~\ref{fig:p1-oam} and \ref{fig:p5-oam}, respectively. In all cases, the rescaling with the appropriate phase correlation length $\xi(p,\ell)$ leads to a universal crosstalk curve for both Kolmogorov turbulence and Gaussian noise, which was once again fitted by Eq.~\eqref{eq:fit}.

\emph{Universality for BG modes.---}
By applying the same scheme to BG modes, we again obtain universal crosstalk curves. 
In this case, we choose to additionally analyze a wider beam waist, in order to expose the intrinsic multi-ring structure of BG beams as presented in Fig.~\ref{fig:bgmodes}.
In particular, we consider a set of BG modes with the same beam waist $w_0$ (see Figs. \ref{fig:p1-lg} and \ref{fig:p1-oam}) as previously chosen for the LG modes, as well as a set of BG modes with a five times larger waist, i.e., $5w_0$, (see Figs. \ref{fig:bg5-lg} and \ref{fig:bg5-oam}).
In both cases, and for either medium, the rescaled crosstalk collapses onto universal curves. Although their particular shape slightly differs from the universal curves found for LG modes, to facilitate comparison, we decided to retain the same fitting function~\eqref{eq:fit}.


\begin{figure*}[ht]
\centering
\includegraphics[width=\linewidth]{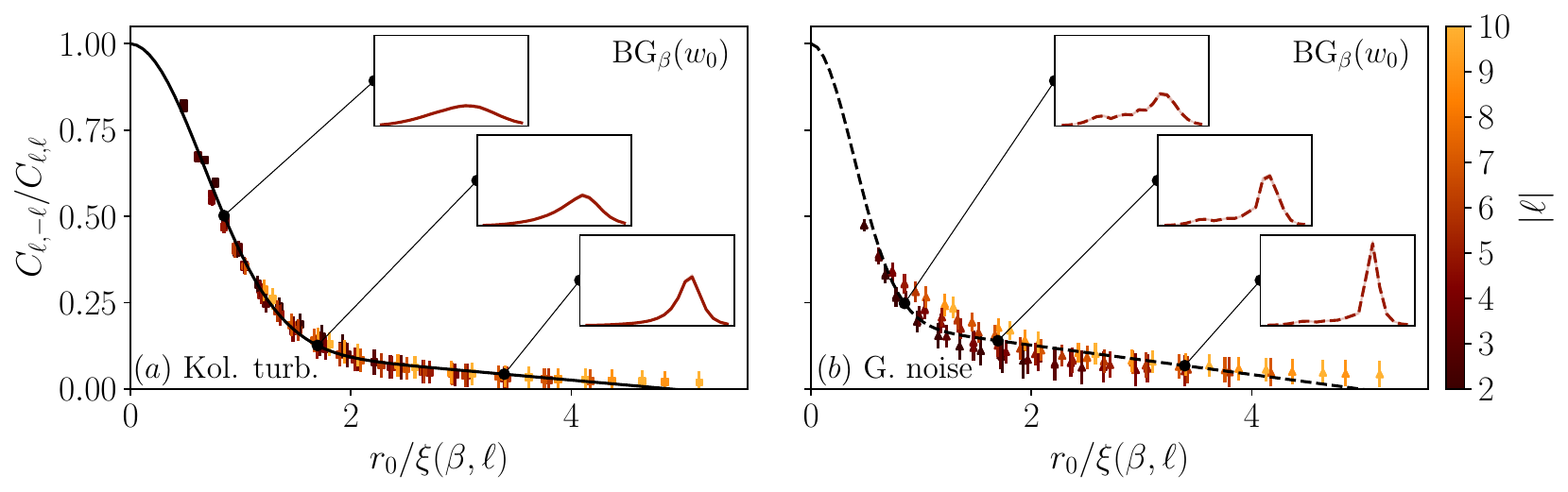}
\caption{Same as in Fig.~\ref{fig:p1-lg}, but for incident BG$_{\beta}^{\ell}(z=0)$ which were projected onto $[$BG$_{\beta}^{\mp\ell}(z=L)]$ at the receiver side, both mode sets were prepared with an increased beam waist of $w_0$. The channel geometry is the same as described in the model section of the main manuscript.}
\label{fig:bg5-lg}
\end{figure*}

\begin{figure*}[ht]
\centering
\includegraphics[width=\linewidth]{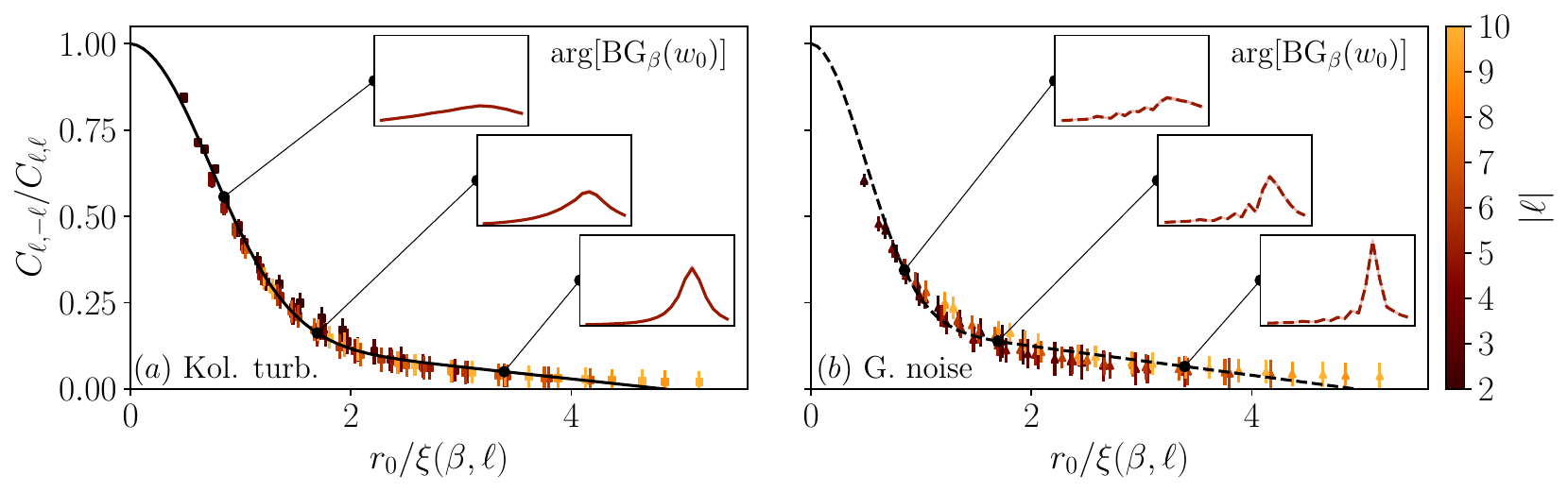}
\caption{Same as in Fig.~\ref{fig:p1-lg}, but for incident BG$_{\beta}^{\ell}(z=0)$ which were projected onto arg$[$BG$_{\beta}^{\mp\ell}(z=L)]$ at the receiver side, both mode sets were prepared with a beam waist of $w_0$. The channel geometry is the same as described in the model section of the main manuscript.}
\label{fig:bg5-lg}
\end{figure*}

\begin{figure*}[ht]
\centering
\includegraphics[width=\linewidth]{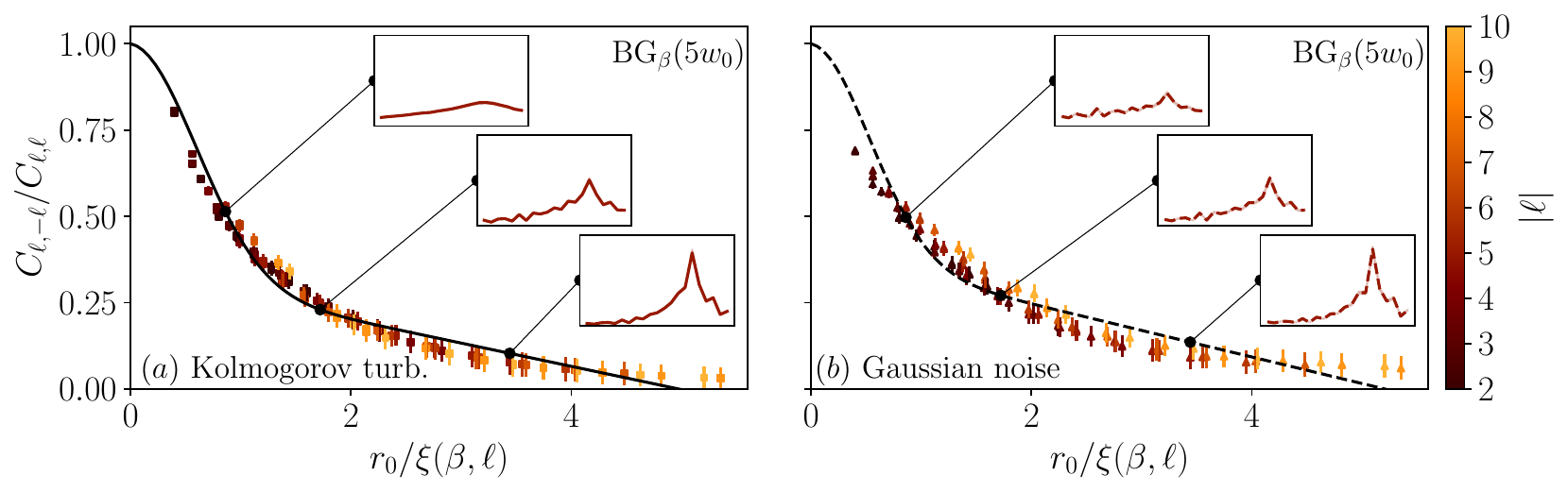}
\caption{Same as in Fig.~\ref{fig:p1-lg}, but for incident BG$_{\beta}^{\ell}(z=0)$ which were projected onto BG$_{\beta}^{\mp\ell}(z=L)$ at the receiver side, both mode sets were prepared with an increased beam waist of $5w_0$. Note that in this case, severeal minor peaks in the crosstalk sections (insets) reflect the modes' mulitring structure (cf. Fig.~\ref{fig:bgmodes}). The channel geometry is the same as described in the model section of the main manuscript but the aperture radius was increased by 30\,\% in order to accommodate these wider modes (at least partially for the largest $|\ell|$).}
\label{fig:bg5-lg}
\end{figure*}

\begin{figure*}[ht]
\centering
\includegraphics[width=\linewidth]{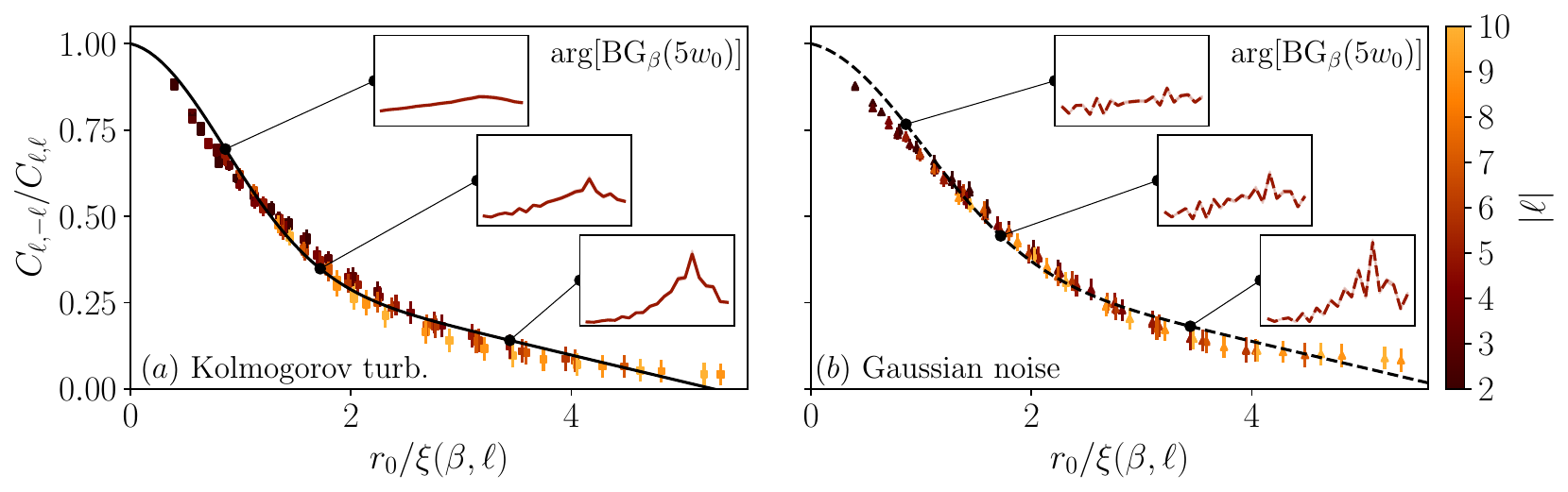}
\caption{Same as in Fig.~\ref{fig:bg5-lg}, but for the incident BG$_{\beta}^{\ell}(z=0)$ projected onto arg$[$BG$_{\beta}^{\mp\ell}(z=L)]$ at the receiver side, both mode sets were prepared with an increased beam waist of $5w_0$. The channel geometry is the same as described in the model section of the main manuscript but the aperture radius was increased by 30\,\%.}

\label{fig:bg5-oam}
\end{figure*}

	\begin{table*}[!htbp]
		\caption{Compilation of all fitting parameters and their errors.}
		\label{tab:paras}
		\centering
		\begin{tabular}{ll|ccc|ccc}
			\toprule
			\multicolumn{2}{c}{\textbf{mode}}  & \multicolumn{3}{c}{\textbf{Kolmogorov turbulence}} & \multicolumn{3}{c}{\textbf{Gaussian noise}}\\ \midrule
			      & & $a$ & $b$ & $c$ &$a$&$b$&$c$ \\  \midrule \midrule
			\multirow{2}{*}{$p=0$\hphantom{xx}}      & LG & 0.208$\,\pm\,$0.014 & 1.687$\,\pm\,$0.012 & 0.032$\,\pm\,$0.004 & 0.129$\,\pm\,$0.009 & 1.644$\,\pm\,$0.007 & 0.007$\,\pm\,$0.002 \\  
			& arg(LG)\hphantom{xx} & \hphantom{x}0.190$\,\pm\,$0.008\hphantom{x} & \hphantom{x}1.621$\,\pm\,$0.007\hphantom{x} & \hphantom{x}0.028$\,\pm\,$0.002\hphantom{x} & \hphantom{x}0.129$\,\pm\,$0.011\hphantom{x} & \hphantom{x}1.567$\,\pm\,$0.009\hphantom{x} & \hphantom{x}0.005$\,\pm\,$0.003\hphantom{x} \\ \midrule
			\multirow{2}{*}{$p=1$}      & LG & 0.335$\,\pm\,$0.010 & 1.456$\,\pm\,$0.009 & 0.063$\,\pm\,$0.004 & 0.249$\,\pm\,$0.013 & 1.449$\,\pm\,$0.011 & 0.037$\,\pm\,$0.004 \\  
			& arg(LG) & 0.291$\,\pm\,$0.022 & 1.597$\,\pm\,$0.018 & 0.051$\,\pm\,$0.006 & 0.166$\,\pm\,$0.011 & 1.620$\,\pm\,$0.008 & 0.017$\,\pm\,$0.003 \\ \midrule
			\multirow{2}{*}{$p=5$}      & LG & 0.583$\,\pm\,$0.018 & 1.106$\,\pm\,$0.025 & 0.141$\,\pm\,$0.006 & 0.525$\,\pm\,$0.025 & 1.126$\,\pm\,$0.029 & 0.120$\,\pm\,$0.006 \\  
			& arg(LG) & 0.464$\,\pm\,$0.031 & 1.491$\,\pm\,$0.028 & 0.104$\,\pm\,$0.009 & 0.322$\,\pm\,$0.019 & 1.544$\,\pm\,$0.014 & 0.060$\,\pm\,$0.006 \\ \midrule \midrule
            \multirow{2}{*}{$w_0$}      & BG & 0.139$\,\pm\,$0.006 & 0.048$\,\pm\,$0.006 & 0.0028$\,\pm\,$0.002 & 0.212$\,\pm\,$0.008 & 0.548$\,\pm\,$0.012 & 0.043$\,\pm\,$0.003 \\  
			& arg(BG) & 0.168$\,\pm\,$0.007 & 1.022$\,\pm\,$0.007 & 0.035$\,\pm\,$0.002 & 0.208$\,\pm\,$0.007 & 0.686$\,\pm\,$0.008 & 0.042$\,\pm\,$0.002 \\ \midrule
            \multirow{2}{*}{$5w_0$}      & BG & 0.335$\,\pm\,$0.010 & 0.847$\,\pm\,$0.015 & 0.067$\,\pm\,$0.003 & 0.400$\,\pm\,$0.013 & 0.754$\,\pm\,$0.021 & 0.008$\,\pm\,$0.004 \\  
			& arg(BG) & 0.402$\,\pm\,$0.014 & 1.201$\,\pm\,$0.019 & 0.076$\,\pm\,$0.004 & 0.438$\,\pm\,$0.017 & 1.440$\,\pm\,$0.075 & 0.075$\,\pm\,$0.005 \\ \bottomrule
		\end{tabular}
	\end{table*}



\end{document}